# Nonlinear electromagnetic response of few-layer graphene: A nonperturbative description.

D.V.Fil

Institute for Single Crystals, National Academy of Sciences of Ukraine, 60 Nauky Avenue, Kharkiv 61072, Ukraine,
V.N. Karazin Kharkiv National University, 4 Svobody Square, Kharkiv 61022, Ukraine

Nonperturbative approach based on exact solution of Boltzmann kinetic equation in the relaxation time approximation is developed for the study of nonlinear response of electron-doped few-layer graphene to a high-frequency electromagnetic field. It is shown that nonperturbative approach can be applied to a two-dimensional conductor with an arbitrary isotropic spectrum of carriers.

## 1. Introduction

Linear dependence of electron spectrum on the momentum causes many unusual properties of graphene. One of them is a strong nonlinear electromagnetic response. Nonlinearity of the response can be seen from the dependence of the velocity on the momentum of charged quasiparticles with the spectrum $\varepsilon(p) = vp$. The velocity is equal to $\mathbf{v}(\mathbf{p}) = d\varepsilon(p)/d\mathbf{p} = v\mathbf{p}/p$. In a very strong electric field $\mathbf{E}(t) = \mathbf{E}_0 \sin\omega t$ the momentum oscillates with the frequency $\omega$: $\mathbf{p} \approx e\mathbf{E}_0 \tau \sin\omega t$, where $\tau$ is the relaxation time ($\omega\tau \ll 1$). Then the electrical current contains all odd Fourier harmonics and it is independent of $E_0$: $j(t) = (4/\pi)env[\sin\omega t + \sin(3\omega t)/3 + \ldots]$, where $n$ is the density of carriers.

In [1,2] nonlinear electromagnetic properties of graphene were studied within the quasiclassical approach based on the Boltzmann kinetic equation. The density matrix approach was developed in [3–6]. It was predicted in [3-6] that the third-harmonic generation (THG) intensity in electron-doped graphene has the main peak at $\hbar\omega = 2\varepsilon_F/3$, where $\varepsilon_F$ is the Fermi energy, and two minor peaks at $\hbar\omega = \varepsilon_F$ and $\hbar\omega = 2\varepsilon_F$. Similar results were obtained within the diagrammatic approach [7]. Strong enhancement of THG in a system of two graphenes, one of which is the electron-doped, and the other, the hole-doped, was predicted in [8]. The approaches used in [1-8] are perturbative ones. In [9] a nonperturbative theory of nonlinear electromagnetic response of graphene was developed. The theory is based on the exact solution of the kinetic Boltzmann equation within the relaxation-time approximation. One of interesting results of [9] is the absence of optical bistability in graphene predicted in [10] but not confirmed in [9]. Note that in [10] the exact solution of the kinetic Boltzmann equation was also used but the answer was expanded in series in the amplitude of the electric field. The difference between the semiclassical [1,2,9,10] and quantum [3-8] approaches is that the former ones take into account only the electron band. The graphene spectrum near Dirac points contains the electron and hole bands that touch each other in Dirac points. A nonperturbative approach that takes into account two bands was developed in [11,12]. The results of [12] are based on a heuristic solution of the time-dependent Dirac equation for the two-component wave function (graphene Bloch equation). It was found in [12] that the contribution of the hole band into the nonlinear response of electron-doped graphene is essential only at $\hbar\omega > \varepsilon_F$. While the approach [9] is not valid at high frequencies (for instance, it cannot reproduce peaks in THG intensity [3-8]), at low frequencies it gives adequate description. In particular, in [9] the power-induced transparency in graphene and third-, fifth- and seventh-harmonic generation at large input power were described quantitatively.

The electron spectrum of bilayer graphene differs from one of monolayer graphene. The bilayer graphene has AB stacking of layers and the spectrum of the electron band can be approximated by free electron spectrum $\varepsilon(p) = p^2/2m$, where $m$ is the effective mass. Therefore at $\hbar\omega \ll \varepsilon_F$ and low temperature the electron-doped bilayer graphene should not demonstrate any nonlinear electromagnetic response. In the general case nonlinear response of bilayer graphene can be comparable with one of monolayer graphene [8,13]. The spectrum of few-layer graphene depends on stacking of layers. Quite interesting situation is realized in ABC stacked *N*-layer graphenes. Their low-energy electron spectrum can be approximated as [14,15]

$$\varepsilon_N(p) = v_0^N p^N / t_\perp^{N-1}, \qquad (1)$$

where $v_0 \simeq 10^8$ cm/s is the Fermi velocity in the monolayer graphene, and $t_\perp \approx 0.4$ eV is the nearest-neighbor interlayer hopping energy. Note that Eq. (1) gives the spectrum of the monolayer and bilayer graphene as well. One can expect that the spectrum (1) with $N > 2$ reveals itself in an unusual nonlinear electromagnetic response. In this paper we study nonlinear response of graphene with the spectrum (1) using the semiclassical nonperpurbative approach. The main attention is given to the four-layer graphene ($N = 4$) for which an analytical expression for the nonlinear part of the electrical current can be obtained. For $N = 4$ we calculate the dependence of the transmission, reflection and adsorption coefficients and the efficiency of the third-harmonic generation on the intensity of the incident wave. The results are compared with ones for monolayer graphene.

**2. Nonperturbative expression for the electrical current**

We start from the Boltzmann kinetic equation in the relaxation time approximation

$$\frac{\partial f}{\partial t} + eE(t)\frac{\partial f}{\partial p_x} = -\frac{f - f_0}{\tau}, \qquad (2)$$

where $f_0 = f_0(p_x, p_y)$ is the equilibrium distribution function. The electric field is directed along the $x$-axis. The exact solution of Eq. (2) with the initial condition $f(p_x, p_y, t)|_{t=0} = f_0$ has the form [16]

$$f(\mathbf{p}, t) = e^{-\frac{t}{\tau}} f_0(p_x - p_E(t,0), p_y) \\ + \frac{1}{\tau} \int_0^t dt' e^{-\frac{t-t'}{\tau}} f_0(p_x - p_E(t,t'), p_y), \qquad (3)$$

where

$$p_E(t,t') = e \int_{t'}^t dt'' E(t''). \qquad (4)$$

At $t \gg \tau$ the terms proportional to $\exp(-t/\tau)$ can be neglected and Eq. (3) reduces to

$$f(\mathbf{p}, t) = \int_0^\infty d\xi e^{-\xi} f_0(p_x - p_E(t, t - \xi\tau), p_y). \qquad (5)$$

The electrical current induced by the field $E(t)$ is calculated as

$$j(t) = \frac{g_s g_v e}{(2\pi\hbar)^2} \\ \times \int_0^\infty d\xi e^{-\xi} \int d\mathbf{p} \frac{d\varepsilon(p)}{dp_x} f_0(p_x - p_E(t, t - \xi\tau), p_y), \qquad (6)$$

where $g_s$ and $g_v$ are the spin and valley degeneracy (for graphene $g_s = g_v = 2$). We specify the case of low temperature $T$ ($k_B T \ll \varepsilon_F$, where $k_B$ - is the Boltzmann constant) and take $f_0$ in the form of a step function. Then the integration over $p_x$ in Eq. (6) can be done analytically and we obtain

$$j(t) = \frac{g_s g_v e p_F}{2(\pi\hbar)^2} \int_0^\infty d\xi e^{-\xi} \int_0^1 dy [\varepsilon(p_+[t,\xi,y]) \qquad (7)$$
$$-\varepsilon(p_-[t,\xi,y])],$$

where

$$p_\pm(t,\xi,y) = p_F \sqrt{1 + \Pi_{t,\xi}^2 \pm 2\Pi_{t,\xi}\sqrt{1-y^2}},$$

$\Pi_{t,\xi} = p_E(t, t-\xi\tau)/p_F$ and $p_F$ is the Fermi momentum. Equation (7) gives the current for an arbitrary field $E(t)$ and for an arbitrary isotropic spectrum $\varepsilon(p)$.

For a monochromatic field $E(t) = E_0 \sin \omega t$ and small $E_0$ Eq. (7) reduces to the Drude formula

$$j(t) = \frac{e^2 n \tau v_F}{p_F} E_0 \frac{\sin \omega t - \omega\tau \cos \omega t}{1 + (\omega\tau)^2}, \qquad (8)$$

where $v_F = v(p_F) = (d\varepsilon(p)/dp)|_{p=p_F}$ is the Fermi velocity, and $n$ is a two-dimensional density of carriers ($n = g_s g_v p_F^2 / 4\pi\hbar^2$). The condition of smallness of $E_0$ in the pure limit $\omega\tau \gg 1$ is given by the inequality $eE_0/\omega p_F \ll 1$, and in the dirty limit $\omega\tau \ll 1$, by the inequality $eE_0\tau/p_F \ll 1$. At large $E_0$ the main term in the current (7) in the system with the spectrum (1) is proportional to $E_0^{N-1}$. In the dirty limit it is equal to

$$j(t) = env_F(N-1)! \left(\frac{eE_0\tau}{p_F}\right)^{N-1} \sin\omega t |\sin\omega t|^{N-2}. \qquad (9)$$

Up to the factor $(N-1)!$ Eq. (9) coincides with the estimate $j(t) = env[p(t)]$ with $p(t) = eE(t)\tau$.

## 3. Transmission, reflection and absorption of high-intensity incident wave in a four-layer graphene

At even $N$ one can obtain from Eq. (7) the explicit dependence of the current on the electric field. In particular, for $N=4$

$$j(t) = env_F \int_0^\infty d\xi e^{-\xi} \left(\Pi_{t,\xi} + \Pi_{t,\xi}^3\right). \qquad (10)$$

The current (10) can be presented as a sum of a linear and a cubic in $E_0$ terms: $j = j^{(1)} + j^{(3)}$. The linear term is equal to

$$j^{(1)}(t) = env_F \tilde{E}_\omega [A_1(\omega\tau)\sin\omega t \qquad (11)$$
$$+ B_1(\omega\tau)\cos\omega t],$$

where $\tilde{E}_\omega = eE_0/\omega p_F$, $A_1(x) = x/(1+x^2)$ and $B_1(x) = -x^2/(1+x^2)$. Note that for $N=4$ the exact expression (11) coincides with the approximate one (8). The nonlinear part of the current reads

$$j^{(3)}(t) = env_F \left(\tilde{E}_\omega\right)^3 [A_{(3,1)}(\omega\tau)\sin\omega t$$
$$+ B_{(3,1)}(\omega\tau)\cos\omega t + A_{(3,3)}(\omega\tau)\sin(3\omega t) + \qquad (12)$$
$$B_{(3,3)}(\omega\tau)\cos(3\omega t)]$$

with

$$A_{(3,1)}(x) = \frac{9}{2}\frac{x^3}{(1+x^2)(1+4x^2)},$$

$$B_{(3,1)}(x) = -\frac{9x^4}{(1+x^2)(1+4x^2)},$$

$$A_{(3,3)}(x) = -\frac{3}{2}\frac{x^3(1-11x^2)}{(1+x^2)(1+4x^2)(1+9x^2)}, \qquad (13)$$

$$B_{(3,3)}(x) = \frac{9x^4(1-x^2)}{(1+x^2)(1+4x^2)(1+9x^2)}.$$

The obtained dependences (11) and (12) allows to calculate the transmission, reflection and absorption coefficients for a monochromatic wave as a function of the incident intensity. We consider the normal incidence. We are interested in a frequency range in which the wavelength is much larger than the thickness of four-layer graphene. In this case the graphene can be treated as a zero-thickness boundary between the upper and the lower half-spaces. We take the electric field of the transmitted wave in the form

$$E_{tr}(t) = E_{tr}\sin(\omega t + kz),$$

the field of the incident wave, in the form

$$E_{inc}(t) = E_{inc}^a \sin(\omega t + kz) + E_{inc}^b \cos(\omega t + kz),$$

and the field of the reflected, in the form

$$E_{ref}(t) = E_{ref}^a \sin(\omega t - kz) + E_{ref}^b \cos(\omega t - kz).$$

The dielectric constant of the environment is assumed to be equal to unity. The boundary conditions yield the following relations between the amplitudes

$$E_{inc}^a = E_{tr}\left[1 + A_1\beta_\omega + A_{(3,1)}\beta_\omega\left(\frac{eE_{tr}}{p_F\omega}\right)^2\right],$$

$$E_{inc}^b = \beta_\omega E_{tr}\left[B_1 + B_{(3,1)}\beta_\omega\left(\frac{eE_{tr}}{p_F\omega}\right)^2\right], \qquad (14)$$

$$E_{ref}^a = E_{tr} - E_{inc}^a, \qquad E_{ref}^b = -E_{inc}^b,$$

where the coefficients $A_1, B_1, A_{(3,1)}, B_{(3,1)}$ are given by Eqs. (13) at $x = \omega\tau$ and

$$\beta_\omega = \frac{2\pi e^2 n v_F}{\omega c p_F}$$

($c$ is the light velocity). From Eqs. (14) we obtain the system of equations for the intensities of the incident ($I_{inc}$), transmitted ($I_{tr}$) and reflected ($I_{ref}$) waves:

$$\tilde{I}_{inc} = \tilde{I}_{tr}[(1 + A_1\beta_\omega + A_{(3,1)}\beta_\omega\tilde{I}_{tr})^2$$
$$+ \beta_\omega^2(B_1 + B_{(3,1)}\tilde{I}_{tr})^2],$$
$$\tilde{I}_{ref} = \tilde{I}_{tr}\beta_\omega^2[(A_1 + A_{(3,1)}\tilde{I}_{tr})^2 \qquad (15)$$
$$+ (B_1 + B_{(3,1)}\tilde{I}_{tr})^2],$$

where the intensities are normalized to the quantity

$$I_\omega = \frac{c}{8\pi}\left(\frac{\omega p_F}{e}\right)^2$$

($\tilde{I}_{inc(ref,tr)} = I_{inc(ref,tr)} / I_\omega$). In the pure limit ($\omega\tau \gg 1$) Eqs. (15) reduce to

$$\tilde{I}_{inc} = \tilde{I}_{tr}\left[1 + \beta_\omega^2\left(1 + \frac{9}{4}\tilde{I}_{tr}\right)^2\right],$$

$$\tilde{I}_{ref} = \tilde{I}_{inc} - \tilde{I}_{tr}.$$

(16)

In the dirty limit ($\omega\tau \ll 1$) we introduce the variables $\tilde{I}^\tau_{inc(ref,tr)} = I_{inc(ref,tr)} / I_\tau$, where

$$I_\tau = \frac{c}{8\pi}\left(\frac{p_F}{e\tau}\right)^2$$

and reduce Eqs. (15) to the form

$$\tilde{I}^\tau_{inc} = \tilde{I}^\tau_{tr}\left(1 + \beta_\tau + \frac{9}{2}\beta_\tau \tilde{I}^\tau_{tr}\right)^2,$$

$$\tilde{I}^\tau_{ref} = \tilde{I}^\tau_{tr}\beta_\tau^2\left(1 + \frac{9}{2}\tilde{I}^\tau_{tr}\right)^2,$$

(17)

where

$$\beta_\tau == \frac{2\pi e^2 n v_F \tau}{c p_F}.$$

Solving Eqs. (15) (or, Eqs.(16),(17) in the corresponding limiting cases) we find the coefficients of transmission $T = I_{tr}/I_{inc}$, reflection $R = I_{ref}/I_{inc}$ and adsorption $A = 1 - R - T$. We emphasize that Eqs. (15) do not take into account losses of energy caused by generation of third harmonic. We will see below that these losses are small.

At low intensities the transmission and reflection are determined by $\beta_\omega$ or $\beta_\tau$: In the pure limit $T = 1/(1+\beta_\omega^2)$ and $R = \beta_\omega^2/(1+\beta_\omega^2)$. In the dirty limit $T = 1/(1+\beta_\tau)^2$, $R = \beta_\tau^2/(1+\beta_\tau)^2$ and $A = 2\beta_\tau/(1+\beta_\tau)^2$. At high input intensity the transmission decreases by the law $T \propto 1/I_{inc}^{2/3}$ and it goes to zero at infinite intensity. Using Eq. (9) one can show that in three-layer graphene the transmission coefficient decreases by the law $T \propto 1/I_{inc}^{1/2}$ at large intensity of the incident wave.

**4. Third-harmonic generation in four-layer graphene**

Nonlinear response causes generation of harmonics. In systems with inversion symmetry only odd harmonics are generated. The harmonics are generated by the component the electrical current that oscillates with the corresponding frequency. Usually the intensity of the third harmonic is the largest one. To calculate the efficiency of THG (the ratio of the intensity of the third harmonic $I^{(3)}$ to the input intensity $I_{inc}$) we should obtain the equation for the electric field of the wave emitted at the frequency $3\omega$. This field is taken in the form

$$E_\pm^{(3)} = \pm[E_g^a \sin(3\omega t \pm 3kz) + E_g^b \cos(3\omega t \pm 3kz)],$$

where the upper/lower sign corresponds to the field in the upper(lower) half-space. The boundary conditions yield

$$(1+A_1\beta_\omega)E_g^a - B_1\beta_\omega E_g^b =$$
$$= -A_{(3,3)}\beta_\omega E_{tr}\left(\frac{eE_{tr}}{p_F\omega}\right)^2,$$
$$B_1\beta_\omega E_g^a + (1+A_1\beta_\omega)E_g^b =$$
$$= -B_{(3,3)}\beta_\omega E_{tr}\left(\frac{eE_{tr}}{p_F\omega}\right),$$
(18)

where $A_{(3,3)}$ and $B_{(3,3)}$ are given by Eqs. (13) at $x = \omega\tau$. In Eqs. (18) we neglect the terms of higher orders in $E_{tr}$ and $E_g^{a,b}$. Using Eqs. (18) we find $E_g^a$ and $E_g^b$ and calculate the efficiency of THG:

$$G = \frac{I^{(3)}}{I_{inc}} = T^3\left(\frac{I_{inc}}{I_\omega}\right)^2 \frac{\beta_\omega^2(A_{(3,3)}^2 + B_{(3,3)}^2)}{(1+A_1\beta_\omega)^2 + \beta_\omega^2 B_1^2}.$$
(19)

Note that in Eq. (19) the transmission coefficient $T$ depends on the input intensity $I_{inc}$. In the pure limit Eq. (19) reduces to

$$G = \frac{T^3}{16}\left(\frac{I_{inc}}{I_\omega}\right)^2 \frac{\beta_\omega^2}{1+\beta_\omega^2}.$$
(20)

It the dirty limit we obtain

$$G = \frac{9T^3}{4}\left(\frac{I_{inc}}{I_\tau}\right)^2 \frac{\beta_\tau^2}{(1+\beta_\tau)^2}.$$
(21)

At small input intensity the efficiency $G$ is proportional to the second power of the intensity of the incident wave. At high input intensity $T \propto 1/I_{inc}^{2/3}$ and $G$ approaches the constant value. In the pure limit this value is equal to

$$G_{max}^{pure} = \frac{1}{81(1+\beta_\omega^2)}.$$
(22)

and in the dirty limit, to

$$G_{max}^{dirty} = \frac{1}{9(1+\beta_\tau)^2}.$$
(23)

One can see that even at very large intensity of the incident wave relative losses caused by THG do not exceed 0.02 in the pure limit and 0.2 in the dirty limit (we take into account the emission into the upper and the lower half-spaces).

It is instructive to compare the efficiency of THG in monolayer and four-layer graphene. Let us do it in the dirty limit. In this limit the current (7) in the monolayer graphene is given by the following Fourier series

$$j(t) = env_0[f_1\left(\frac{eE_{tr}\tau}{p_F}\right)\sin\omega t$$
$$+ f_3\left(\frac{eE_{tr}\tau}{p_F}\right)\sin(3\omega t) + ...],$$
(24)

where the functions $f_n(x) = \frac{1}{\pi}\int_0^{2\pi} dy \sin(ny) g[x\sin(y)]$ are Fourier coefficients. The function $g$ is defined as

$$g(z) = \frac{2}{\pi} \int_0^\infty d\xi \, e^{-\xi} \int_0^1 dy [\sqrt{1 + z^2\xi^2 + 2z\xi\sqrt{1-y^2}}$$
$$-\sqrt{1 + z^2\xi^2 - 2z\xi\sqrt{1-y^2}}\,].$$

Using the relations (14) we obtain the equation for the intensities of the incident and transmitted waves

$$\tilde{I}_{inc}^\tau = \left[ \sqrt{\tilde{I}_{tr}^\tau} + \beta_\tau f_1\left(\sqrt{\tilde{I}_{tr}^\tau}\right) \right]^2 \tag{25}$$

from which we calculate the dependence of $\tilde{I}_{tr}^\tau$ on $\tilde{I}_{inc}^\tau$. Then using the boundary conditions for the third harmonic we obtain the equation for $I^{(3)}$:

$$\sqrt{\tilde{I}_\tau^{(3)}} + \beta_\tau f_1\left(\sqrt{\tilde{I}_\tau^{(3)}}\right) = \beta_\tau f_3\left(\sqrt{\tilde{I}_{tr}^\tau}\right), \tag{26}$$

where $\tilde{I}_\tau^{(3)} = I^{(3)}/I_\tau^{(0)}$. Finally we find the dependence $I^{(3)}(\tilde{I}_{inc}^\tau)$ and calculate $G = I^{(3)}/I_{inc}$ as a function of $\tilde{I}_{inc}^\tau$.

## 5. Numerical estimates and discussion

To estimate the values of the effects described we consider a four-layer graphene with the density of carries $n = 4 \cdot 10^{12}$ cm$^{-2}$. It corresponds to $\varepsilon_F \approx 46$ meV. For such $\varepsilon_F$ the spectrum (1) is a good approximation. Taking the frequency $\omega = 2\pi \cdot 10^{12}$ c$^{-1}$ we obtain $\beta_\omega \approx 0.65$ and $I_\omega \approx 0.28$ MW/cm$^2$. Considering the dirty limit we take $\tau = 0.1$ ps. For such $\tau$ we get $\beta_\tau \approx 0.41$ and $I_\tau \approx 0.72$ MW/cm$^2$. To compare the results for four-layer and monolayer graphene we specify the density of carries in the monolayer graphene $n = 10^{12}$ cm$^{-2}$ (the same density per layer). It corresponds to $\varepsilon_F \approx 0.12$ eV. Taking $\tau = 0.1$ ps we obtain $\beta_\tau \approx 0.26$ and $I_\tau \approx 0.18$ MW/cm$^2$.

In Fig.1 we present the dependence of the transmission and reflection coefficients on the intensity of the incident wave for four-layer graphene in the pure limit calculated for the parameters given above.

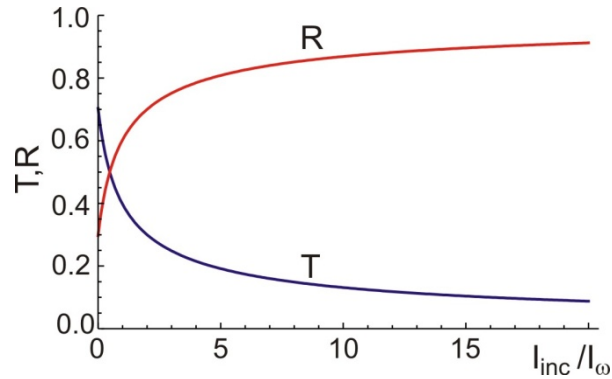

Fig 1. Transmission and reflection coefficients in four-layer graphene in the pure limit versus the intensity of the incident wave.

In Fig.2 the transmission, reflection and absorption coefficients in a dirty four-layer graphene as functions of the intensity of the incident waves are shown. Note that in physical units (MW/cm$^2$) the ranges of input intensities in Fig. 1 and Fig. 2 are almost the same. One can see that the main difference between the pure and dirty limits is nonzero absorption in the latter case. The absorption coefficient depends non-monotonically on the intensity of the incident wave and decreases at large input power. The transmission decreases and the reflection increases at large input power in the pure and in the dirty limits.

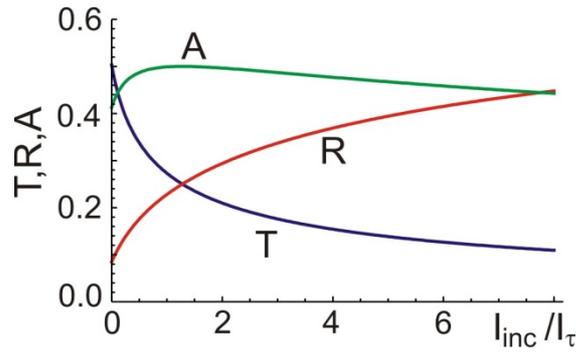

Fig 2. Transmission, reflection and absorption coefficients in four-layer graphene in the dirty limit versus the intensity of the incident wave.

In Fig. 3 the efficiency of THG in pure and dirty four-layer graphenes is presented. One can see that quadratic dependence of the efficiency on the intensity of the incident wave survives only at very small intensities. For the parameters considered the maximum efficiency does not exceed few percents. At the same input power the efficiency of THG in a dirty graphene is larger than in a pure one. The efficiency of THG is the increase function of the intensity of the incident wave.

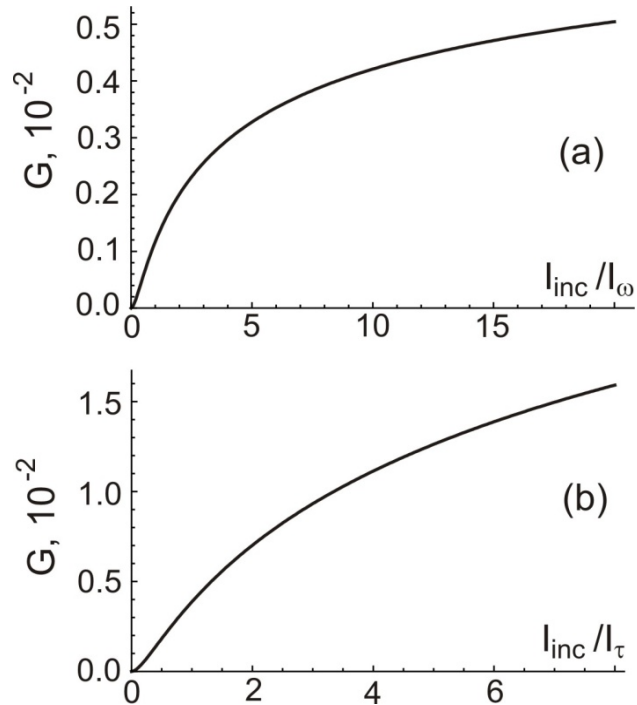

Fig 3. The efficiency of third-harmonic generation in the pure (a) and the dirty (b) limit in four-layer graphene versus the intensity of the incident wave.

To compare the behavior of four-layer and monolayer graphene we calculate the transmission, reflection, and absorption coefficients in a dirty monolayer graphene for the same $\tau$ and for the same density of carries per layer. The result is displayed in Fig. 4. Note that for the parameters specified the reference intensity $I_\tau$ for monolayer graphene is in four times smaller than for four-layer graphene. In physical units the range of intensities in Fig. 4 is in two times smaller than in Figs. 1-3. Figs. 2 and 4 illustrate the difference in the nonlinear response of few-layer and monolayer graphene. The former one demonstrates the power-induced reflectance while the latter one, the power-induced transparency. We emphasize that this difference emerges if the frequency of the incident wave satisfies the condition $\omega < \varepsilon_F / \hbar$ that corresponds to the frequency range up to several terahertz.

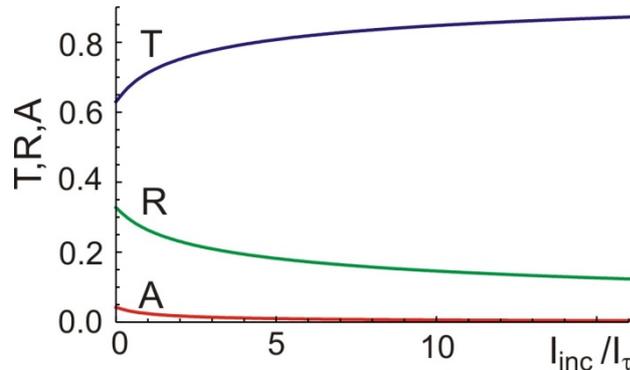

Fig. 4. Transmission, reflection and absorption coefficients in monolayer graphene in the dirty limit versus the intensity of the incident wave.

The efficiency of THG of monolayer graphene is shown in Fig. 5. One can see that it is in two orders smaller than one for the four-layer graphene (Fig. 3b). Another difference is that the efficiency Fig. 5 is a nonmonotonic function of the intensity of the incident wave with the maximum at rather small intensities. The differences are connected with that the amplitude of the third harmonic of the electrical current (24) saturates at large electric field, while the amplitude of the third harmonic of the current (12) is proportional to the third power of the electric field. The saturation of THG efficiency in Fig. 3 is solely due to decrease of the transmission $T$.

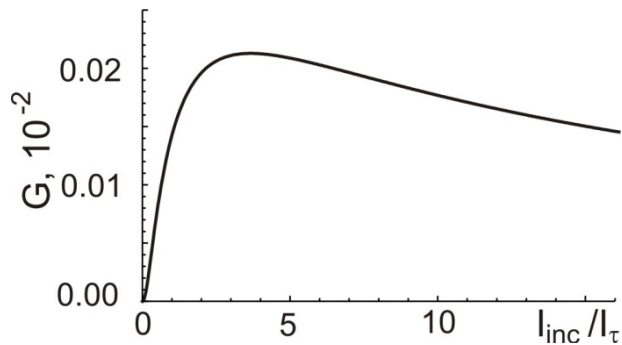

Fig 5. The efficiency of third-harmonic generation in the dirty limit in monolayer graphene versus the intensity of the incident wave.

One should note that in this study we neglect mixing of the first harmonic with generated third harmonic. This effect results in fifth-harmonic generation, a reduction of reflection and lowering of the efficiency of third-harmonic generation at very large input intensities. In the range of $I_{inc}$ considered two latter effects are small, less than one percent.

## 6. Conclusion

In conclusion, we have shown that nonperturbative theory of nonlinear electromagnetic response developed for monolayer graphene can be generalized to the system with an arbitraty isotropic spectrum of carriers. The nonperturbative approach is applied to the study of nonlinear electromagnetic properties of few-layer graphenes with ABC stacking of layers. It is established that nonlinear behavior of electron-doped *N*-layer graphenes with N>2 differs significantly from one of monolayer graphene. We predict that the transmission coefficient $T$ of three- and four-layer graphene decreases under increase in the intensity of the incident wave $I_{inc}$ by the asymptotic law $T \propto 1/I_{inc}^{(N-2)/(N-1)}$. We show that the efficiency of third-harmonic generation in four-layer graphene approaches the constant value at high $I_{inc}$, in contrast to the case of monolayer graphene where this quantity reaches its maximum at certain $I_{inc}$ and then decreases under further increase of $I_{inc}$. The difference is connected with saturation of the amplitude of third harmonic of an electrical current induced in monolayer graphene by a monochromatic field and the absence of such saturation in few-layer graphene. The efficiency of third-harmonic generation in a four-layer graphene can reach several percents which is two orders in magnitude larger than the maximum efficiency in monolayer graphene.

**7. Note added**

The analysis in Sec. 5 does not take into account the deviation of the spectrum of four-layer graphene from the dependence Eq. (1). The use of the spectrum (1) is justified if $\varepsilon_F \ll t_\perp$. In addition, one should require that $\varepsilon\left(p_F(1+\max(\Pi_{t,\xi}))\right) \ll t_\perp$. For the parameters considered in Sec. 5 the first condition is satisfied, but the second condition is satisfied only at rather small intensity of the incident wave. The analysis that takes into account the deviation of the spectrum from Eq. (1) shows that in a four-layer graphene the power-induced reflectance takes place only at small input intensity, while at large input intensity the four-layer graphene behaves similar to a monolayer one: it demonstrates the power-induced transparency. Due to the same reason the THG efficiency in a four layer graphene is similar to one of monolayer graphene at large input intensity.